\documentclass[10pt]{article}
\usepackage{amsmath}
\usepackage{amssymb}
\usepackage[mathscr]{eucal}
\usepackage{cases}
\usepackage{graphicx, subfigure,}
\usepackage{caption}
\usepackage{graphicx}
\renewcommand{\paragraph}{\roman{paragraph}}

\newtheorem{theorem}{\scshape \mdseries  Theorem}[section]
\newtheorem{lemma}[theorem]{\scshape \mdseries  Lemma}
\newtheorem{coro}[theorem]{\scshape \mdseries  Corollary}

\begin{document}

\title{\bf Symmetric Toeplitz-Structured Compressed Sensing Matrices\thanks{
Supported by National Natural Science Foundation of China (11071002), Program for New Century Excellent
Talents in University (NCET-10-0001), Key Project of Chinese Ministry of Education
(210091), Specialized Research Fund for the Doctoral Program of
Higher Education (20103401110002), Scientific Research Fund for Fostering Distinguished Young Scholars of Anhui
University(KJJQ1001).}
}
\author{Tao Huang, Yi-Zheng Fan\thanks{Corresponding author.
 E-mail addresses: fanyz@ahu.edu.cn(Y.-Z. Fan), huangtaooo@hotmail.com (T. Huang), zhu\_m@163.com (M. Zhu)}, \ Ming Zhu\\
  {\small  \it $1.$ Key Laboratory of Intelligent Computing and Signal Processing of Ministry of Education,}\\
{\small \it Anhui University, Hefei 230039, P. R. China} \\
  {\small  \it $2.$ School of Mathematical Sciences, Anhui University, Hefei 230601, P. R. China} \\
 }
\date{}
\maketitle

\noindent
{\bf Abstract}\  How to construct a suitable measurement matrix  is still an open question in compressed  sensing.
A significant part of the recent work is that the measurement matrices are not completely random on the entries but exhibit considerable structure.
In this paper, we proved that the symmetric Toeplitz matrix and its transforms can be used as measurement matrix and recovery signal with high probability.
Compared with random matrices (e.g. Gaussian and Bernullio matrices) and some structured matrices (e.g. Toeplitz and circulant matrices),
we need to generate fewer independent entries to obtain the measurement matrix while the effectiveness of recovery does not get worse.
Furthermore, the signal can be recovered more efficiently by the algorithm.



\newpage
\indent

\section{Introduction}

 Compressed sensing has a number of potential applications in image processing, geophysics, medical imaging, computer science as well as other areas of science and technology.
Due to \cite{donoho4} by Donoho and  \cite{candes} by Candes, Romberg and Tao, this area makes great progress after 2006.
Compressed sensing focus on recovering $x$ from knowledge of $y=Ax$, where $A$ is a suitable  $k\times n$ measurement matrix (also called {\it CS matrix}) and $k\ll n$.
As we know, when the signal $x=(x_i)_{i=1}^n\in \mathbb{R}^n$ is {\it $m$-sparse} (i.e. the number of non-zero coefficients of the vector $x$ is at most $m$)
  and the CS matrix $A$ holds certain conditions such as {\it restricted isometry property} (RIP)\cite{can} for all $m$-sparse vectors $x$, the solution $x^*$  recovers $x$ exactly.  Actually, recovering $x$ can be solved by $l_1$-minimization instead of $l_0$-minimization:
$$\min\|x\|_1     \mbox{~~subject to~~}    y=Ax\eqno(1.1)$$
where the $l_p$-norm is defined  $\|x\|_p=(\sum_{j=1}^n |x_j|^{p})^{1/p} $, as usual.

As a weaker version of  RIP, {\it RIP of order $3m$ }can be expressed as follows.
 Let $T\subset\{1,2,\ldots,n\}$ and $A_T$ be the $k\times|T|$ submatrix obtained by retaining the columns of $A$ corresponding to the indices in $T$;
 then, there exists a constant $\delta_{3m}\in(0,1/3)$ such that
$$(1-\delta_{3m})\|x\|_2^2\leq\|A_Tx\|_2^2\leq (1+\delta_{3m})\|x\|_2^2\eqno(1.2)$$
holds for all subsets $T$ with $|T| \le 3m$.
It was shown that random matrices with entries drawn independently from certain probability distributions satisfy RIP of order $3m$ with high probability for
every $\delta_{3m} \in (0, 1/3)$ provided $k \ge const \cdot m \ln(n/m)$ \cite{candes3,can1}.
We refer to such matrices as independent and identically distributed CS matrices or simply {\it IID CS matrices}.

The problem that how to choose a suitable CS matrix is a main field in compressed sensing.
It had been proved that Gaussian or Bernoulli  matrices can be used as IID CS matrices \cite{candes3,rau3,rude,bar,candes}
if we are allowed to choose the entries of sensing matrix freely.
But most applications do not allow a free choice of the entries of the sensing matrix and enforce a particularly structure on the matrix.
 Recently, the random Toeplitz or circulant matrices introduced in \cite{raz,hol,rau} are estimated as CS matrix,
 where the entries of the vector generating the Toeplitz or circulant matrices are chosen at random according to a suitable probability distribution, which then allowed for providing recovery guarantees for $l_1$-minimization.
 Compared to Bernoulli or Gaussian matrices, random Toepliz and circulant matrices have the advantages that they require $O(n)$ instead of $n^2$ random entries to be generated.  In\cite{Fan,Fan2} Fan et.al show that some symmetric random (not necessarily IID) matrices satisfied RIP.

  Motivated by the work \cite{raz}, we  show that if a probability distribution $P(a)$ yields an IID CS matrix (having unit-norm columns in expectation)
  then  a  $k\times n$ partial {\it symmetric} Toeplitz matrix $A$ (also having unit-norm columns in expectation) of the form
$$
\left[
\begin{array}{ccccc}
 a_n & a_{n-1} &\cdots& a_2& a_1 \\
 a_{n-1} & a_n &\cdots& a_3 & a_2 \\
 \cdots & \cdots & \ddots & \ddots & \vdots \\
 a_{n-k+1} & a_{n-k+2} &\cdots &\cdots & a_k
\end{array}
\right],\eqno(1.3)
$$
where the entries $\{a_i\}_{i=1}^{n}$ have been drawn independently from $P(a)$,
is also a measurement matrix in the sense that it satisfies RIP of order $3m$ with high probability for every $\delta_{3m}\in(0,1/3)$  provided $k \geq const\cdot
m^3\cdot \ln(n/m) $.
Compared with Toeplitz matrix in \cite{raz}, we only need to generate $n$ independent entries rather than $n+k-1$ independent entries.
So, $k-1=O(\ln n)$ less entries are reduced to form CS matrices, while the effective of recovery does not get worse.

\section{Main Result}
The proof of the main result adopts the technique used in \cite{raz} with a slight modification.
 We use the celebrated Hajnal-Szemer\'edi theorem on equitable coloring of graphs \cite{haj} to partition a $k \times |T|$ partial symmetric Toeplitz-structured matrix $A_T$ into roughly $O(m^2)$ IID submatrices with dimensions approximately equal to $O(k/m^2) \times |T|$.
We also extend the main result to some variants of symmetric Toeplitz matrices.

\begin{lemma}
For fixed $n,m$, let $P(a)$ be a probability
distribution that generates a $k \times n$ IID matrix having unit-norm
columns in expectation such that, for every $\delta_{3m}\in(0,1/3)$ and every
$T\subset\{1,2,\ldots,n\}$ with $|T|=3m$, the $k \times |T|$ IID submatrix
obtained by retaining the columns corresponding to the indices in $T$ satisfies (1.2) with probability at least $$1-e^{-f(k,m,\delta_{3m})},\eqno(2.1)$$
where $f(k,m,\delta_{3m})$ is some real-valued function of $k, m \mbox{ and } \delta_{3m}$.

Let $\{a_i\}_{i=1}^{n}$ be a sequence of random variables drawn independently
from the same distribution, and $A$ be a $k\times n$  symmetric Toeplitz matrix of the form given in (1.3). Then, for every $\delta_{3m}\in(0,1/3)$ and every $T\subset\{1,2,\ldots,n\}$ with $|T|=3m$, the symmetric Toeplitz submatrix $A_T$ satisfies (1.2) with probability at least
$$1-e^{-f(\lfloor k/q\rfloor,m,\delta_{3m})+ln(q)},\eqno(2.2)$$
where $q=3m(6m-1)+1$.
\end{lemma}

{\bf Remark 1.} Let the probability distribution $P(a)$ be given by
$$\mathcal{N}(0,1/k), ~
\begin{cases}
+\sqrt{1/k}       & \mbox{ with probability 1/2},\\
-\sqrt{1/k}        & \mbox{ with probability 1/2},
 \end{cases}
 ~\mbox{or} ~
 \begin{cases}
+\sqrt{3/k}        & \mbox{ with probability 1/6},\\
 ~~   0        & ~\mbox{with probability 2/3},\\
-\sqrt{3/k}        & ~\mbox{with probability 1/6}.
 \end{cases}
\eqno(2.3)$$
Then $$f(k, m, \delta_{3m})=c_0k-3m\ln(12/\delta_{3m})-\ln2,\eqno(2.4)$$ where $c_0 =c_0
(\delta_{3m}) =\delta_{3m}^2/16-\delta_{3m}^3/48$  (see \cite {bar}).

\vspace{3mm}

{\it Proof.} Fix $\delta_{3m}\in(0,1/3)$ and $T\subset\{1,2,\ldots,n\}$ with $|T|=3m$.
Let $A_{T,i}$ denote the $i$-th row of $A_T$ for each $i=1,2,\ldots, k$.
Construct an undirected  graph  $G= (V, E)$ with $V=\{1,2,\ldots,k\}$ and $E=\{(i,i^{'})\in V\times V : i\neq i^{'}, A_{T,i} \mbox { and} A_{T,i^{'}
} \mbox{ are dependent}\}$.
Notice  the  properties of symmetric Toeplitz matrices, $A_{T,i}$ can at most be dependent with
$|T|(2|T|-1)=:q-1$,
which implies that the maximum degree $\Delta$ of $G$ is given by $\Delta\leq q-1$.
By Hajnal-Szemer\'edi theorem on equitable coloring of graphs \cite{haj}, we
can always partition $G$ using $q$ colors such that
$$\lfloor k/q\rfloor \leq \min_{j \in \{1,2,\ldots,n\}}|C_j|\leq \max_{j \in \{1,2,\ldots,n\}}|C_j|\leq\lceil k/q \rceil,$$
where $\{C_j\}_{j=1}^q$ correspond to the different color classes.

Next, let $A_T^j$ be the $|C_j| \times |T|$ partition submatrix obtained by retaining the
rows of $A_T$ corresponding to the indices in $C_j$, which is an IID submatrix of $A_T$.
Observe that for any $x \in \mathbb{R}^{|T|}$,
$$||A_Tx||_2^2= \sum_{j=1}^q||A_T^jx||_2^2=\sum_{j=1}^q \frac{|C_j|}{k}||\tilde{A}_T^jx||_2^2,\eqno(2.5)$$
where $\tilde{A}_T^j:=\sqrt{\frac{k}{|C_j|}}A_T^j$ (to ensure unit-norm columns in expectation).
So, each $\tilde{A}_T^j$ is a $|C_j| \times|T|$ submatrix with IID entries from the
distribution $P(a)$ and hence, satisfies (1.2) with probability at least
$$1-e^{-f(|C_j|,m,\delta_{3m})}\geq 1-e^{-f(\lfloor k/q \rfloor,m,\delta_{3m})}.\eqno(2.6) $$

Noting that $\sum_{j=1}^q\frac{|C_j|}{k}=1$, from (2.5) the occurrence of the event
$\{\cap_{j=1}^q\tilde{A}_T^j ~\mbox{satisfies (1.2)}\}$
implies that for any $x\in \mathbb{R}^{|T|}$,
$$\sum_{j=1}^q\frac{|C_j|}{k}(1-\delta_{3m})||x||_2^2\leq \sum_{j=1}^q \frac{|C_j|}{k}||\tilde{A}_T^jx||_2^2
\leq \sum_{j=1}^q \frac{|C_j|}{k}(1+\delta_{3m})||x||_2^2,$$
and hence
$$(1-\delta_{3m})||x||_2^2\leq||A_Tx||_2^2\leq(1+\delta_{3m})||x||_2^2.$$
So, $$\{\cap_{j=1}^q\tilde{A}_T^j ~\mbox{satisfies (1.2)}\} \subset \{A_T ~\mbox{satisfies (1.2)}\}.\eqno(2.7)$$
Consequently, from (2.6) and (2.7) we have
\begin{align*}
\Pr(\{A_T~\mbox{satisfies (1.2}\})& =1-\Pr(\{A_T~\mbox{does not satisfies (1.2)}\})\\
 & \geq 1-\Pr(\{\cup_{j=1}^q \tilde{A}_T^j~\mbox{does not satisfies (1.2)}\})\\
& \geq 1-\sum_{j=1}^q \Pr(\{\tilde{A}_T^j~\mbox{does not satisfies (1.2)}\})\\
& \geq1-\sum_{j=1}^q e^{-f(\lfloor k/q \rfloor,m,\delta_{3m})}\\
&=1-e^{-f(\lfloor k/q \rfloor,m,\delta_{3m})+\ln q}.
\end{align*}
This completes the proof of the lemma.\hfill $\blacksquare$

\begin{theorem}
Suppose that $n, m$ are given, and let A be a $k \times n$ partially
symmetric Toeplitz matrix of the form given in (1.3), where the entries $\{a_i\}_{i=1}^{n}$
are drawn independently from one of the probability distributions
given in (2.3). Then, there exist constants $c_1, c_2 > 0$ depending only on $\delta_{3m}$ such that for any $k > c_1 m^3  \ln(n/m) $, $A$ satisfies RIP of
order $3m$ for every $\delta_{3m}\in(0,1/3)$ with probability at least
$$1-e^{-c_2k/m^2}.\eqno(2.8)$$
\end{theorem}

{\it Proof.} For given $\delta_{3m}\in (0,1/3)$, from Lemma 2.1, $A$ satisfies (1.2)
for any $T\subset\{1,2,\cdots,n\}$ of cardinality $3m$ with probability at least
$$1-e^{-c_0\lfloor k/q \rfloor+3m\ln(12/\delta_{3m})+\ln 2+\ln q}\geq 1-e^{-c_0 k/{18m^2} +3m\ln(12/\delta_{3m})+\ln 2+\ln({18m^2})+c_0},$$
and there are ${n \choose 3m}\leq(en/3m)^{3m}$ such subsets. Consequently,
union bounding over these subsets yields that $A$ satisfies RIP of order $3m$ with probability at least
$$ 1-e^{-c_0 k/{18m^2} +3m[\ln(12/\delta_{3m})+\ln(n/3m)+1]+\ln 2+\ln({18m^2})+c_0}.\eqno(2.9)$$
Next, fix $c_2 >0$ and pick $c_1 >54c_3/(c_0 -18c_2)$, where $c_3 =\ln(12/\delta_{3m})+2\ln 2+c_0+4$.
Then, for any $k\geq c_1m^3\ln(n/m)$, the exponent in the exponential in (2.9) is upper bounded by $-c_2k/m^2$
and this completes the proof of the theorem.\hfill $\blacksquare$


By a similar discussion as in Theorem 2.2, we have the following two corollaries immediately.

\begin{coro}
The results of Theorem 2.2 apply equally well to the left-shifted Toeplitz matrix of the form
$$
\left[
\begin{array}{ccccc}
 a_1 & a_2 & \cdots & a_{n-1} & a_n \\
 a_2 & a_3 & \cdots & a_n & a_{n-1} \\
 \vdots & / & /  & \vdots  & \vdots\\
 a_k & \cdots &\cdots & a_{n-k+2} & a_{n-k+1}
\end{array}
\right].
$$
\end{coro}

\begin{coro}
Let $\Theta \subset \{1,2,\ldots,n\}$ with $|\Theta|=k$.
Then the submatrix $A_\Theta$ obtained from a symmetric Toeplitz or left-shifted symmetric Toeplitz matrix
by retaining the rows corresponding to the indices in $\Theta$, satisfies Theorem 2.2.
\end{coro}


\section{Applications}
We now describe how the results for symmetric Toeplitz CS matrices lend themselves to (i) identification of LTI systems having sparse impulse responses; and
(ii) recovery of signals that either piecewise constant (PWC) or sparse in the Harr wavelet domain.

\subsection{System Identification}
The application of Toeplitz CS matrices in identifying a LTI system with finite sparse impulse responses has been discussed in \cite{raz}.
Here we discuss the symmetric Toeplitz CS matrices in the identification of such system.
Let $x[0],x[1],\ldots,x[n-1]$ be $m$-sparse impulse responses of a LTI system (of duration $n$).
Let $a[0],a[1],\ldots,a[n-2],a[n-1],a[n-2],\ldots,a[n-k]$ be a sequence of duration $(n + k- 1)$
  which are symmetric to the time $n-1$ (i.e. $a[n-1+t]=a[n-1-t]$ for $t=1,2,\ldots,k-1$), where $a[0],a[1],\ldots,a[n-1]$ have been drawn from one of the probability distributions given in (2.3).
  So, $k-1=O(\ln n)$ less entries are reduced to be generated.
  Then, probing the given system with $a[l]$ yields $y[l] = a[l]\ast x[l]$ and the theory of CS along with Theorem 2.2 guarantees that, with high probability, $x[l]$ can be exactly recovered by solving the convex program
$$x[l]=\arg\left(\underset{z \in \mathbb{R}^n}\min||z||_1,\mbox{~subject to~} y = Az\right), $$
where
$y=
\left[
\begin{array}{c}
y[n-1]\\
y[n] \\
 \vdots \\
y[n+k-2]
\end{array}
\right]
$ and
$
A=\left[
\begin{array}{ccccc}
 a[n-1] & a[n-2] &\cdots& a[1] & a[0] \\
 a[n-2] & a[n-1] &\cdots& a[2] & a[1] \\
 \vdots & \vdots & \ddots & \ddots & \vdots\\
 a[n-k] & a[n-k-1]&\cdots& \cdots & a[k-1]
\end{array}
\right].
$

\subsection{Beyond Sparse Signals}
When signals are sparse in some transform domain $\Psi \neq I$, i.e., $x =\Psi\theta$ and $\theta\in \mathbb{R}^n$ is $m$-sparse,
we need guarantee the product matrix $ A\Psi$ satisfies RIP of order $3m$ for successful recovery of $\theta$.
It's unquestionable when $A$ is a random matrix and $\Psi $ is any orthonormal basis \cite{bar}.
As described in \cite{raz}, the Toeplitz matrices can still be used as CS matrices for some {\it fixed} transformations, even though they lack the universality property because of their highly structured nature.
Here we still use the example in \cite{raz} for recovery of PWC signals by symmetric Toeplitz matrices.
We benefit from this kind of matrices such as only $n$ independent random variables ($k-1=O(\ln n)$ less random variables than Toeplitz matrices), and faster acquisition and reconstruction algorithms.

For example, let $x$ be a $m$-piece PWC signal with form $x = L\theta$, where $\theta\in \mathbb{R}^n$ is $m$-sparse, and $L \in \mathbb{R}^{n\times n}$
is the matrix whose entries are $0$'s above the diagonal and $1$'s everywhere else.
Further, let $\{a_i\}_{i=1}^n$ be a sequence of random variables drawn independently from a distribution that yields an IID CS matrix,
and $A_L \in \mathbb{R}^{k\times n}$ be a cascade of a $k \times n$ symmetric Toeplitz matrix $A$ and the
$n \times n$ differencing operator $D$ (see \cite{raz}).
Then,  $A_L=AD$ has only $n$ degrees of freedom,
 and the product matrix $A_LL = ADL = A$ is a symmetric Toeplitz CS matrix  satisfied RIP with high probability.

\section{Experiment}
Now we compare the performances between the Toeplitz matrices (Toeplitz) and symmetric Toeplitz matrices (S-Toeplitz). Let $x$ be a  discrete signal with length $512$ and  the sparsity $k=20$ whose nonzero entries are $1$ or $-1$.
The classical convex optimization algorithm $\ell_1$-minimization is used for reconstruction. The results of $100$ experiments are summarized in Fig. 4.1.

 \begin{figure}[!h]
  \centering
  \renewcommand\thefigure{\arabic{section}.\arabic{figure}}
    \begin{minipage}[]{.8\textwidth}
      \centering
     \includegraphics[width=0.9\textwidth]{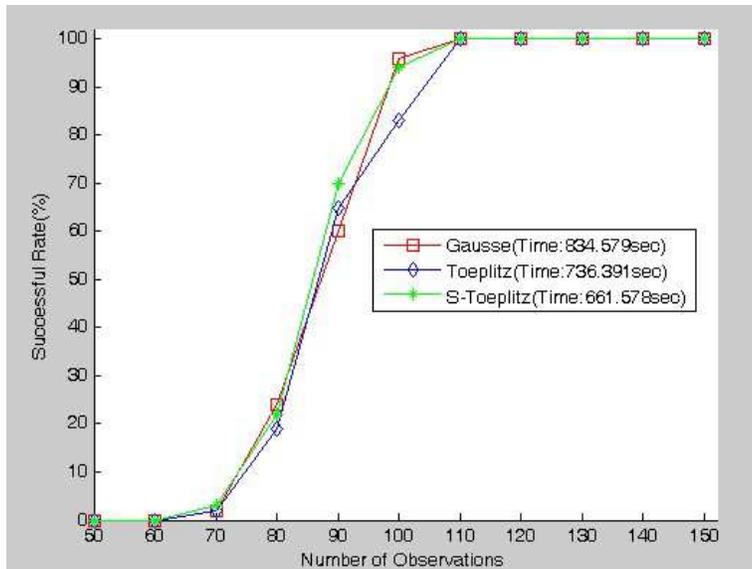}\\
    \caption{Successful rate as a function of measurement number}
   \end{minipage}
     \hspace{0in}
\end{figure}

 Next, we compare the performances through the real image reconstruction experiment.
   The original image is shown in Fig. 4.2, with size of $64\times 64$ and sparsity $m=739$.
   Set measurement number $k=2400$.
   The mean square error (MSE) is defined as $\mbox{MSE}=\frac{\|X-M\|_F}{\|M\|_F}$,
   where $\|\cdot\|_F$ is the Frobenius norm, $X$ is the reconstruction and $M$ is the original image.
The experimental results show that symmetric Toeplitz matrices are suitable measurement matrices.
\begin{figure}[!h]
  \centering
  \renewcommand\thefigure{\arabic{section}.\arabic{figure}}
  \subfigure{
    \begin{minipage}{0.31\textwidth}
      \centering
     \includegraphics[width=0.9\textwidth]{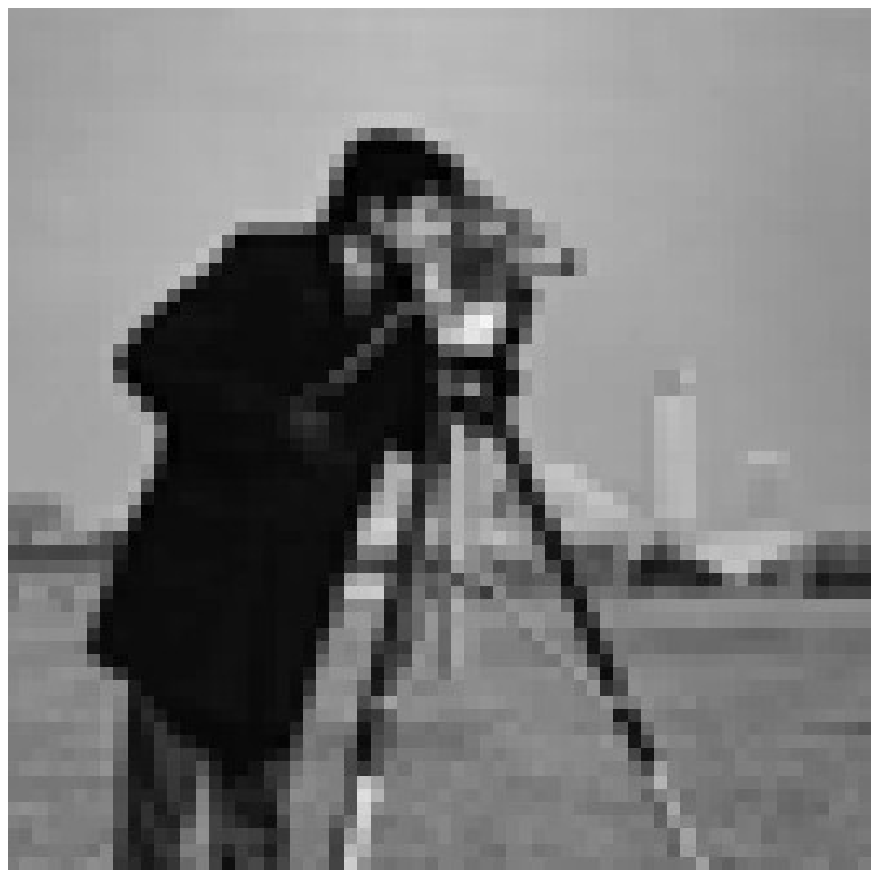}\\
   \caption*{Original image\\}
  \end{minipage}}
     \hspace{0in}
  \subfigure{
    \begin{minipage}{0.31\textwidth}
      \centering
     \includegraphics[width=0.9\textwidth]{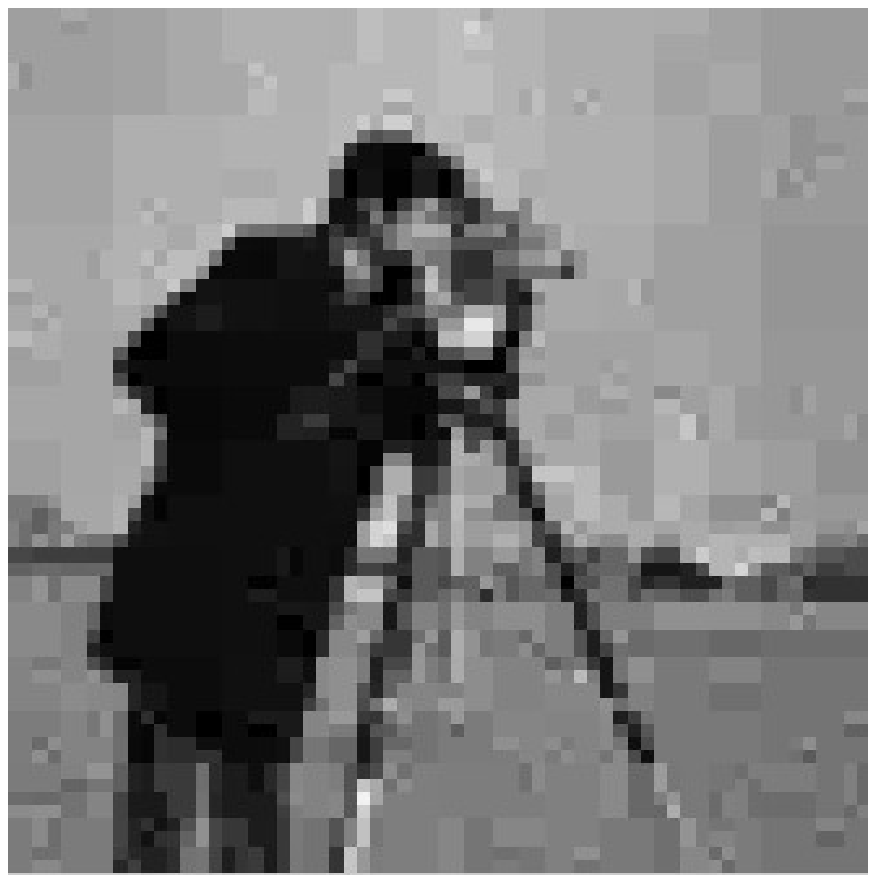}\\
    \caption*{Toeplitz (MSE=0.0668)}
   \end{minipage}}
     \hspace{0in}
       \subfigure{
   \begin{minipage}{0.31\textwidth}
      \centering
     \includegraphics[width=0.9\textwidth]{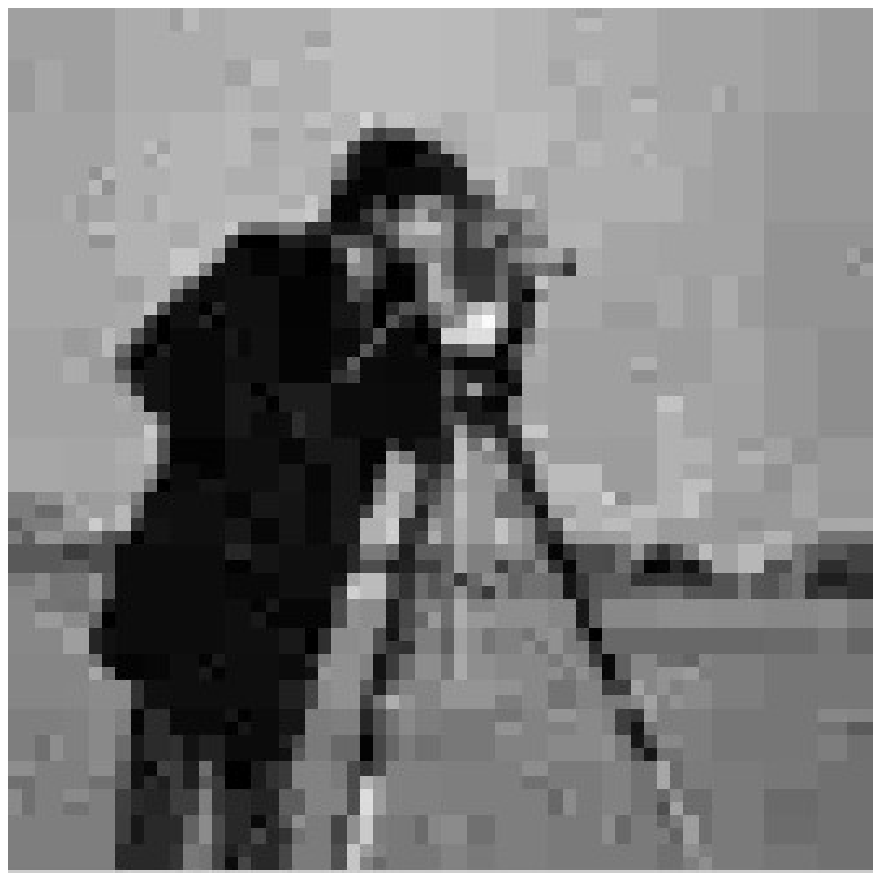}\\
    \caption*{S-Toeplitz (MSE=0.0688)}
     \end{minipage}}
     \hspace{0in}
     \caption{ Real world data reconstruction}
\end{figure}
\section{Conclusion}
We show that symmetric Toeplitz-structured matrices and its transforms  are also sufficient to recover undersampled sparse signals. It provide an desirable alternative of measurement matrix  for a number of application areas because it greatly reduces the computational and storage complexity in large dimensional problems.

\end{document}